\documentclass[useAMS,usegraphicx,usenatbib]{mn2e}
\usepackage{times}
\usepackage{url}
\usepackage{soul}
\usepackage[T1]{fontenc}
\usepackage{ulem}

\usepackage{color}
\usepackage[usenames,dvipsnames]{xcolor}

\title[MLS110213:022733+130617: A new eclipsing polar above the period gap]{MLS110213:022733+130617: A new eclipsing polar above the period gap\thanks{Based on observations made at the Observat\'{o}rio do Pico dos Dias, Brazil, operated by the Laborat\'{o}rio Nacional de Astrof\'{i}sica.}\thanks{Based on observations obtained at the Southern Astrophysical Research (SOAR) telescope, which is a joint project of the Minist\'{e}rio da Ci\^{e}ncia, Tecnologia e Inova\c{c}\~{a}o (MCTI) da Rep\'ublica Federativa do Brasil, the U.S. National Optical Astronomy Observatory (NOAO), the University of North Carolina at Chapel Hill (UNC) and Michigan State University (MSU).}}
\author[Silva et al. 2013]
{K.~M.~G. Silva$^{1,2}$\thanks{E-mail:karleyne@gmail.com},
C.~V. Rodrigues$^{2}$, A. S. Oliveira$^{1}$, L. A. Almeida$^{2,3,4}$, D. Cieslinski$^{2}$
\newauthor
J.~E.~R. Costa$^{2}$ \& F. J. Jablonski$^{2}$\\
$^{1}$ Universidade do Vale do Para\'{i}ba (Univap) -- Av. Shishima Hifumi, 2911 - S\~ao Jos\'e dos Campos - SP -- Brazil.\\
$^{2}$ Instituto Nacional de Pesquisas Espaciais (INPE/MCTI) --
Av. dos Astronautas, 1758 - S\~ao Jos\'e dos Campos - SP -- Brazil\\
$^{3}$ Instituto de Astronomia, Geof\'isica e Ci\^encias Atmosf\'ericas (IAG/USP) -- Rua do Mat\~ao 1226 - S\~ao Paulo -SP -- Brazil \\
$^{4}$ Department of Physics \& Astronomy, Johns Hopkins University, Bloomberg Center for Physics and Astronomy, Room 520, 3400 N Charles St}
\begin{document}

\date{\today}

\pagerange{\pageref{firstpage}--\pageref{lastpage}} \pubyear{2015}

\maketitle

%\label{firstpage}
\def\mls{MLS110213:022733+130617\ }
\def\mmls{MLS110213:022733+130617}
\def\ml{MLS110213}

\begin{abstract} 
    This study confirms \mmls\ as a new eclipsing polar. We performed optical spectroscopic, polarimetric and photometric follow-up of this variable source identified by the Catalina Real Time Transient Survey. Using the mid-eclipse times, we estimated an orbital period of 3.787 h, which is above the orbital period gap of the cataclysmic variables. There are nine other known polars with longer orbital periods, and only two of them are eclipsing. We identified high and low-brightness states and high polarization modulated with the orbital period. The spectra are typical of polars, with strong high ionization emission lines and inverted Balmer decrement. The \mbox{He\,{\sc ii}} 4686 \r{A} line is as strong as H$\beta$. We modelled the photometric and polarimetric bright-state light curves using the {\sc cyclops} code. Our modelling suggests an extended emitting region on the WD surface, with a mean temperature of 9~keV and B in the range 18 -- 33, although the possibility that it could be a two-pole accretor cannot yet be ruled out. The WD mass estimated from the shock temperature is 0.67 M$_{\odot}$. The derived parameters are consistent with the eclipse profile. The distance was estimated as 406$\pm$54~pc using the Period-Luminosity-Colours method. \ml\ populates a rare sub-group of polars, near the upper limit of the period distribution, important to understand the evolution of mCVs. 

\end{abstract}


\section[Introduction]{Introduction}
\label{intro}

AM Her, also known as polars, are a sub-class of magnetic Cataclysmic Variables (mCVs): binary systems, composed of a highly magnetised (B~$\sim 7-120$ MG) white dwarf (WD) primary star and a low-mass red dwarf secondary, in which mass is transferred from the secondary to the primary star by Roche lobe overflow. In contrast to non-magnetic CVs, in which the transferred mass spirals in an accretion disc towards the WD, in AM Her stars the accretion occurs through a magnetic accretion column along the magnetic field lines and the rotation of both stars are synchronised with the orbital period. Intermediate polars (IPs) constitute the other class of mCVs. They are not synchronised and may have an accretion disc that is internally truncated by the magnetic field of the WD. The optical and near-infrared radiation from polars is dominated by cyclotron emission and the X-ray radiation is dominated by bremsstrahlung, both from the accretion column near the WD, the post-shock region. %(PSR).
Cyclotron radiation is highly anisotropic and polarized. Therefore  AM Her are the stellar sources with the largest fraction of polarized light in the sky, reaching 50 per cent in optical bands. See \citet{cropper1990} for a review on polars. 

According to the standard model, the evolution of CVs is based on angular momentum loss that decreases the separation between the stars and, consequently, the orbital period. Thus, the systems evolve from longer to shorter periods. There are two known mechanisms to cause the angular momentum loss in CVs: the magnetic braking (MB  - \citealt{verbunt1981}), which dominates in longer period systems and is driven by the outflowing magnetised stellar wind of the secondary; and the emission of gravitational waves (GW), which prevails in the shorter period systems. Most CVs have periods ranging from 1.2 to 6~h, but there is a deficiency in the number of systems with periods between 2 and 3~h, known as the period gap. The period gap is understood as the transition from the MB to the GW mechanisms, considering that the secondary is out of equilibrium in the MB phase due to the mass loss. When the MB halts, the secondary contracts to its equilibrium radius and detaches from the Roche lobe, interrupting the mass transfer and becoming too faint to be detected. The mass transfer is re-established when the GW mechanism shrinks the Roche Lobe back to contact with the secondary, at orbital periods around 2~h \citep[e.g.,][]{Howell2001,webbink2002}. 

The period gap is not so conspicuous in polars as it is in non-magnetic CVs. The presence of a strong magnetic field in the primary could make the MB mechanism less efficient, which is attributed to the trapping of the stellar wind in the combined magnetosphere of both stars in synchronised systems \citep[and references therein]{webbink2002}. Therefore, evolution of polars is slow and the accretion rates are low in comparison with others CVs.

Another difference between polars and non-magnetic CVs is the scarcity of long-period AM Her systems. On the contrary, IPs are predominatly found above the period gap. 
\citet{chanmugam1984} proposed that IPs evolve to polars as the orbital period decreases as result of standard CV evolution. A problem with this scenario is the different distributions of magnetic field intensities in polars and IPs. \citet{cumming2002} proposed that the WD magnetic field can be reduced by accretion, which could redeem that scenario. However, \citet{zhang2009} performed new calculations considering non-spherical accretion and concluded that the magnetic field is not easily modified by accretion.  Presently it is not yet clear what is the evolutionary relation between mCVs \citep[and references therein]{norton2008,pretorius2013}.

The Catalina Real Time Transient Survey \citep[CRTS;][]{drake2009} improved the number of discoveries of faint transient objects. CRTS is composed by three separate surveys - the Catalina Schmidt Survey (CSS), the Mount Lemmon Survey (MLS) and the Siding Spring Survey (SSS) - that repeatedly scan a combined area of 30 000 deg$^2$ of the sky in both hemispheres, up to a limit of 19-21 mag in $V$ band. This survey has already found more than one thousand CV candidates, which are made public on their website\footnote{\url{http://crts.caltech.edu}}. \mls (hereafter \ml) is one of those candidates: it is a Galactic source that is highly variable, ranging from 16.5 mag to 19.5 mag in the CRTS light curve in timescales of days. We selected it, as a target for detailed observational follow-up, from a larger program of spectroscopic classification of mCV candidates (Oliveira et al., in preparation). 

As of June 2014\footnote {update RKcat7.22, 2014, at \url{http://physics.open.ac.uk/RKcat/}}, 114 polars have been identified \citep[hereafter RK catalogue]{ritter2003,ritter2011}. Among them, 30 systems have periods above the period gap and only 5 of them are eclipsing, being \ml\ one of those systems. \ml\ is included in the RK catalogue based on reports in VSNET\footnote{VSNET, at \url{http://www.kusastro.kyoto-u.ac.jp/vsnet/}. The vsnet-alerts 12847, 12855, 12884, 12888 are related to \ml.}. There are only two other known eclipsing polars having longer periods, namely V895 Cen \citep{1997AJ....113.2231H} and V1309 Ori \citep{2001A&A...374..588S}. 
Here we report the results of the photometric, spectroscopic and polarimetric follow-up of \ml\ and the first modelling of the light and polarization curves.   
In Section~2, we present the data. In Section~3, we show the analysis of the light curve, of the polarimetric data and of the spectra. The modelling of the light and polarization curves using the {\sc cyclops} code is presented in Section~4. The eclipse fitting is explained in Section~5. In Section~6 we estimate the distance to \ml. We discuss our results and present our conclusions in Section~7. Preliminary results were presented in \citet{silva2014}.

\section{Observations and data reduction}

Photometric and polarimetric observations of \ml\ were performed at the Observat\'orio Pico dos Dias (OPD - LNA/MCTI), located in southeast Brazil, using the 1.6-m Perkin-Elmer telescope, while the spectroscopic data were obtained on the SOAR Telescope, on Cerro Pachon, Chile. The journal of observations is presented in Table \ref{tabdados}. The data acquisition and reduction are described in the following sections.

\begin{table*}
\begin{center}
\caption{Log of observations of \ml.}
\label{tabdados}
\begin{tabular}{l c c c c c}
\hline
Date          &     Telescope     &   Instrument   & Filter  & Exp. time (s) & Duration (h)          \\
\hline
%2005 Dec 20 - 2013 Feb 01&                                &  CCS  0.6m      &   $V$  &           &  7.2 years$^{1}$      \\
%2005 Dec 25 - 2013 Mar 01&                                &  MLS  1.5m      &   $V$  &           &  7.2 years $^{1}$    \\
2011 August 14     & 1.6 m P-E    & Cam1+IkonL      &  $I_C$ &   60     &  2.7               \\
2011 September 22  & 1.6 m P-E    & Polarimeter + IkonL + $\lambda$/4    & $R_C$  &   90     &    5.0            \\
2011 October 27    & 1.6 m P-E    & Cam1+IkonL      &  $R_C$ &  100     &  5.1      \\
2011 October 28    & 1.6 m P-E    & Cam1+IkonL      &  $R_C$ &   60     &  2.5     \\
2012 August 13     & 1.6 m P-E    & Cam1+IkonL      & $I_C$  &   60     &  3.3              \\
2012 September 10  & 4.1 m SOAR   & Goodman HTS     &   -    &  1200    &  1.0                 \\
2013 August 23     & 1.6 m P-E    & Cam1+IkonL      & Clear  &   10     &  1.5 \\
\hline
\end{tabular}
\end{center}

\end{table*}
 
\subsection{Photometry}

\begin{figure}%% l b r t
\centering
\includegraphics[trim= 3.3cm 1cm 2cm 1cm, clip,width=0.58\textwidth]{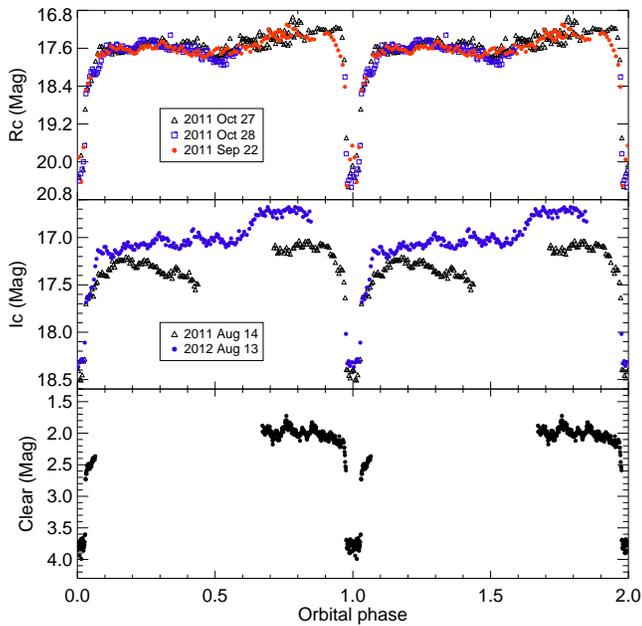}
\caption{
OPD light curves of \ml\, folded with the ephemeris estimated in Section \ref{curvas}. From top to bottom: ${R_C}$ band, ${I_C}$ band and white light. White light magnitudes are not calibrated.} 
\label{fig:lc1}
\end{figure}

Photometric data in $R_C$, ${I_C}$ and white light (clear) bands were obtained employing a thin, back-illuminated 2048$\times$2048 E2V CCD 42-40 mounted in an IkonL camera.
The exposure times were scaled according to the atmospheric conditions, and acurate timings were provided by a GPS receiver. Bias and dome flat-field exposures were used for correction of the detector read-out noise and sensitivity using \textsc{iraf}\footnote {\textsc{iraf} is distributed by the National Optical Astronomy Observatories, which are operated by the Association of Universities for 
Research in Astronomy, Inc., under cooperative agreement with the National Science Foundation.} standard routines. The differential aperture photometry was performed with the \textsc{daophot~ii} package, adopting as reference the star USNO~B1.0~31$-$0032240 (${\rm R}$~=~15.38~mag; ${\rm I}$~=~15.23~mag). The conversion between ${R}$ and Landolt's ${R_C}$ for this object indicates a difference of only 0.03 mag \citep{kidger2003}. The resulting light curves in ${R_C}$, ${I_C}$ and white light are shown in Fig.~\ref{fig:lc1}.

\subsection{Polarimetry}

\begin{figure}%% l b r t
\centering
\includegraphics[trim= 3.1cm 1cm 2cm 1.6cm, clip,width=0.57\textwidth]{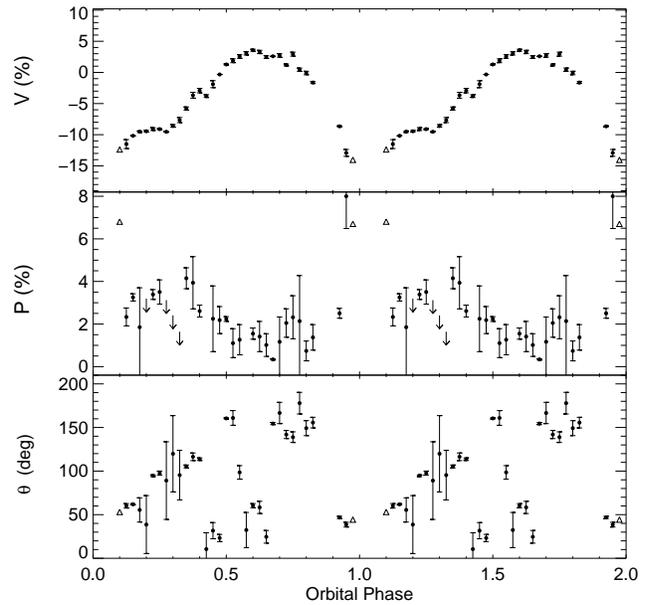}
\caption{Polarimetry of \ml\ in ${R_C}$ band. The data points were grouped in 40 orbital phase bins. The lack of points around phase 0 corresponds to the eclipse in which we had no signal to measure the polarization. From top to bottom, the panels show the circular polarization, linear polarization, and position angle of the linear polarization. The triangles are points with no errors estimation.} 
\label{pol_p}
\end{figure}

We obtained ${R_C}$ polarimetric data using a CCD camera modified by the polarimetric module, as described in \citet{magalhaes1996}. The data reduction followed the standard procedures using {\sc iraf}. The polarization was calculated according to \citet{magalhaes1984} and \citet{rodrigues1998} using the package {\sc pccdpack} \citep{pereyra2000} and a set of IRAF routines developed by our group\footnote{\url{http://www.das.inpe.br/~claudia.rodrigues/polarimetria/reducao_pol.html}}. Each set of 8 images is used to produce one measurement of linear and circular polarizations. We grouped the images in the following way: 1-8, 2-9, 3-10, and so on. Hence the polarization points are not independent measurements. The polarization position angle correction to the equatorial reference system was performed using standard polarized stars. The polarization of the non-polarized standards is consistent with zero, therefore no instrumental polarization correction was applied. We could not determine the correct sign of the circular polarization - if positive or negative. But we could detect changes in this signal. The linear polarization is always a positive value, which introduces a bias into any measurement: the measured polarization value is greater than the true polarization value \citep{simmons1985}. In particular, measurements having $P/\sigma_{P} < 1.4$  provides only upper limits to the real polarization value. The polarization of \ml\ was bias-corrected following \citet{vaillancourt2006}. Figure \ref{pol_p} shows the resulting polarization curves for \ml, binned in 40 orbital phases. The lack of points around phase 0 corresponds to the eclipse in which we had no signal to measure the polarization.

The ordinary and extraordinary counts of the polarimetric data were summed to obtain the total counts of each object. Hence the polarimetric data allowed us to perform differential photometry and to obtain an additional \ml\ light curve, which is also plotted in Fig.~\ref{fig:lc1}. 

\subsection{Spectroscopy}

\begin{figure}%% l b r t
\centering
\includegraphics[clip,width=0.48\textwidth]{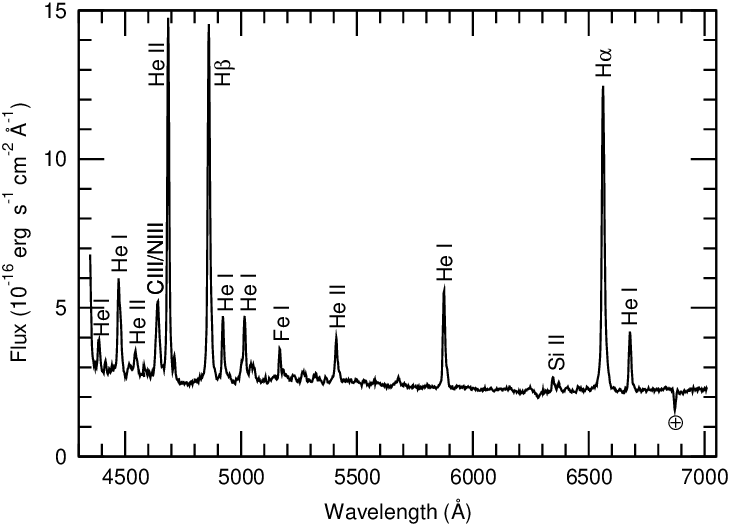}
\caption{Average spectrum of \ml.} 
\label{fig:spec}
\end{figure}

Spectra of \ml\ were obtained in queue mode on September 10, 2012 using the SOAR Telescope, Chile, with the Goodman High Throughput Spectrograph \citep{2004SPIE.5492..331C} and a Fairchild 4096$\times$4096 CCD with 15 micron/pixel. The spectrograph was set to operate with 600 l/mm VPH grating, 1.68 arcsec slit and the GG~385 blocking filter, yielding a spectral resolution of 7~\r{A} FWHM in the range from 4350 to 7005 \r{A}. Three 1200~s exposures were obtained, reaching individual S/N ratio of 45 at the continuum. Quartz lamps calibration flats and bias images were taken to correct the CCD signature, while CuHeAr lamp exposures were used for wavelength calibration, which resulted in typical 0.8 \r{A} or about 45 km s$^{-1}$ calibration RMS residuals. The [\mbox{O\,{\sc i}}] 5577 \r{A} telluric spectral line was used to assess the calibration accuracy. Two spectra of the HR~1544 spectrophotometric standard \citep{Hamuy1992} were used for flux calibration. The data reduction, spectra extraction and calibrations were performed with IRAF standard routines. The average spectrum of \ml~ is presented in Fig.~\ref{fig:spec}. 

\section{Results} 
 
\subsection{Light curves and ephemeris}
\label{curvas}

\begin{figure}%% l b r t
\centering
\includegraphics[trim= 2.1cm 0.8cm 1.5cm 1.5cm, clip,width=0.5\textwidth]{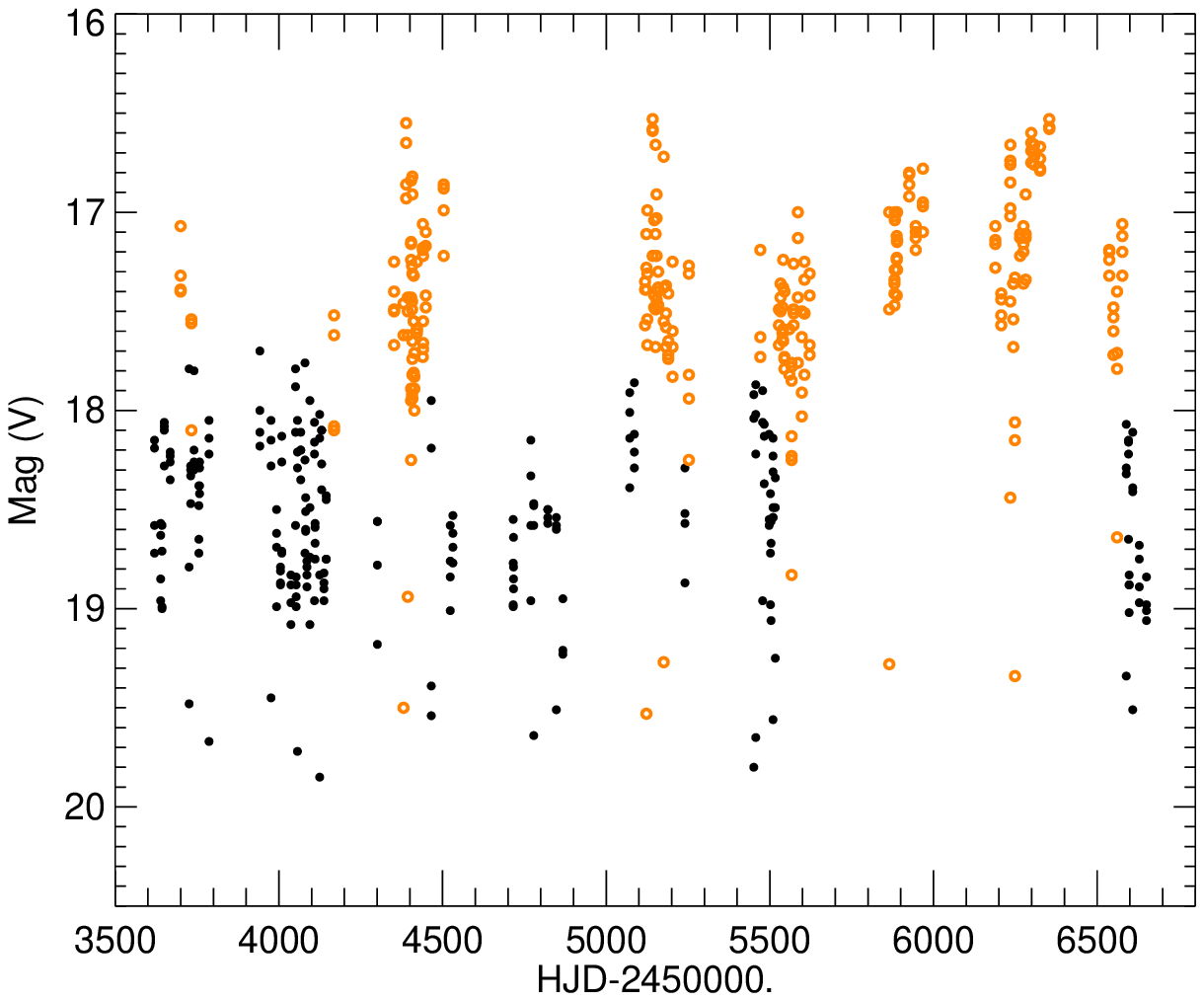}
\includegraphics[trim= 2.1cm 0.8cm 1.5cm 1.5cm, clip,width=0.5\textwidth]{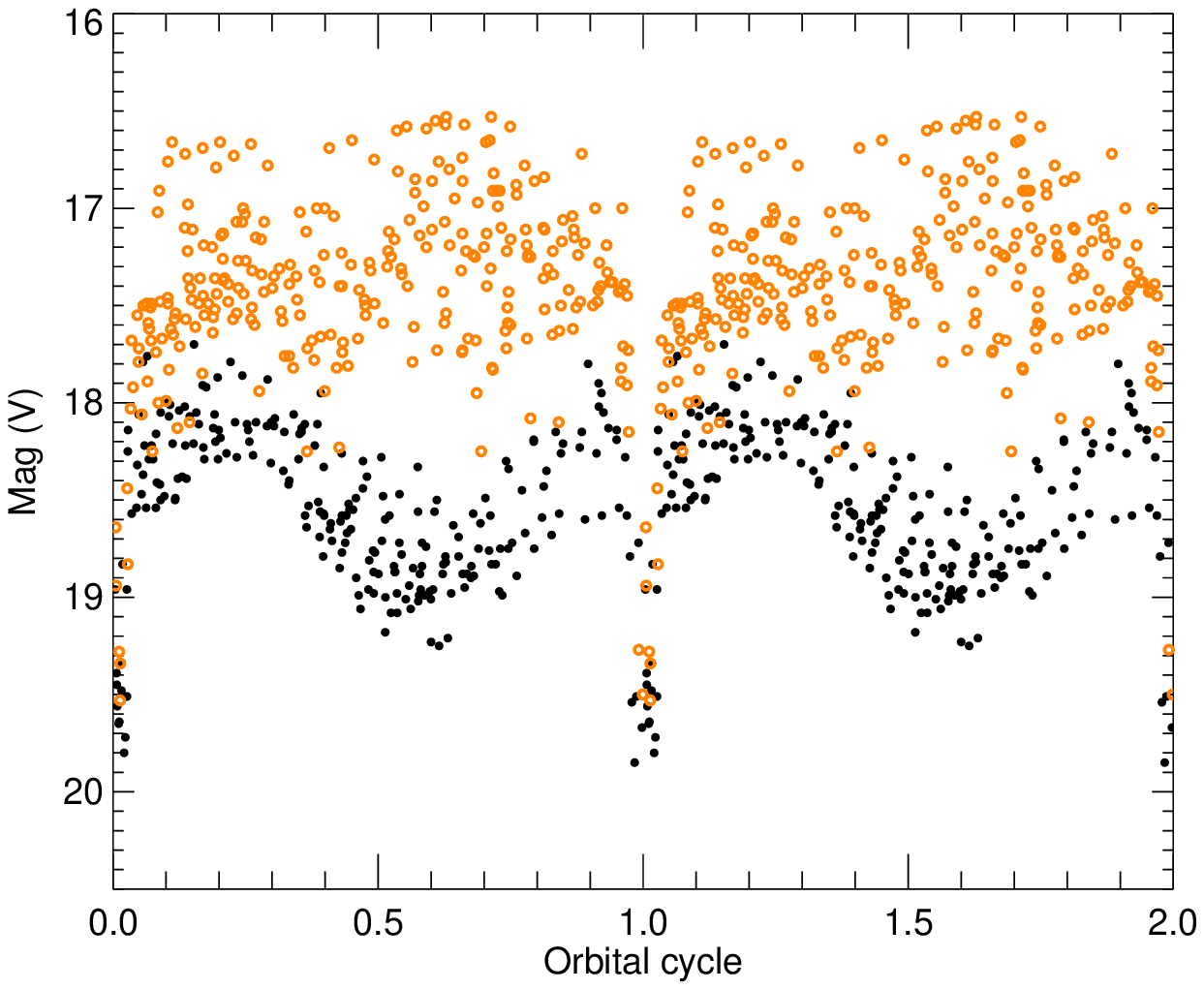}
\includegraphics[trim= 0.65cm 1.8cm 0.3cm 2.6cm, clip,width=0.5\textwidth]{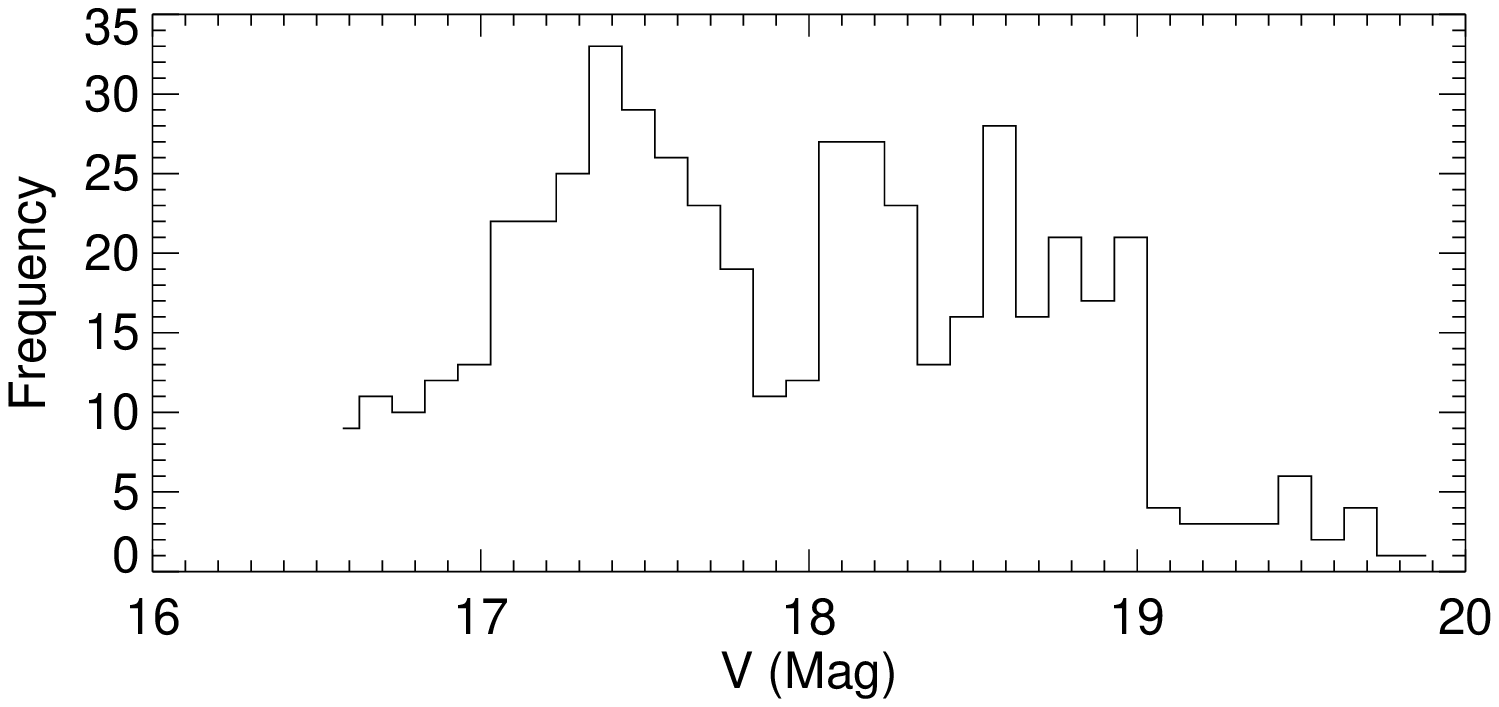}
\caption{Top panel: V band light curves of \ml\ showing the high (open circles) and low (filled circles) brightness states along 8.3 years of CRTS observations. Middle panel: The same data as in the top panel, folded using our ephemeris. Bottom panel: Histogram of the magnitudes of \ml\ using a bin size of 0.1 mag.
} 
\label{fig:lc3}
\end{figure}

We analysed the photometric ${I_C}$, ${R_C}$ and white light data obtained at OPD as well as the $V$ data available from CRTS. This later dataset consists of 513 data points from the CSS and MLS surveys, spanning 8.3 years with a typical cadence of 4 consecutive exposures every visit, which are typically separated by 2-30 days. CRTS data points have magnitudes in the range from 16.5 to 19.5 mag and are shown in Fig.~\ref{fig:lc3}.

\ml\ light curves obtained at OPD have higher temporal resolution than CRTS data and clearly show eclipses, whose amplitude reaches $\sim$3 mag. We determined the timings of the six OPD eclipses as the midpoints of the  bottom-of-eclipse light curves, the uncertainty being half of the exposure plus readout times (Table~\ref{tabephe}).  Our measurements were converted to BJD(TDB) using the online code developed by \citet{eastman2010}. These timings were used to a first estimate of \ml\ ephemeris. We then used the CRTS data to visually refine the ephemeris. The resulting ephemeris is

\begin{equation}
%T_{\rm ecl} (HJD) = 2\,455\,862.6001\;(\pm3) + 0.15779872(\pm6) \times E.\; 
%\end{equation}
%\begin{equation}
 T_{\rm ecl} = (BJD) 2\,455\,862.6002\;(\pm3) + 0.15779878(\pm6) \ \times E.\ 
\label{eq-ephem}
\end{equation}

\begin{figure}%% l b r t
\centering
\includegraphics[trim= 3.5cm 1cm 2.0cm 1.75cm, clip,width=0.58\textwidth]{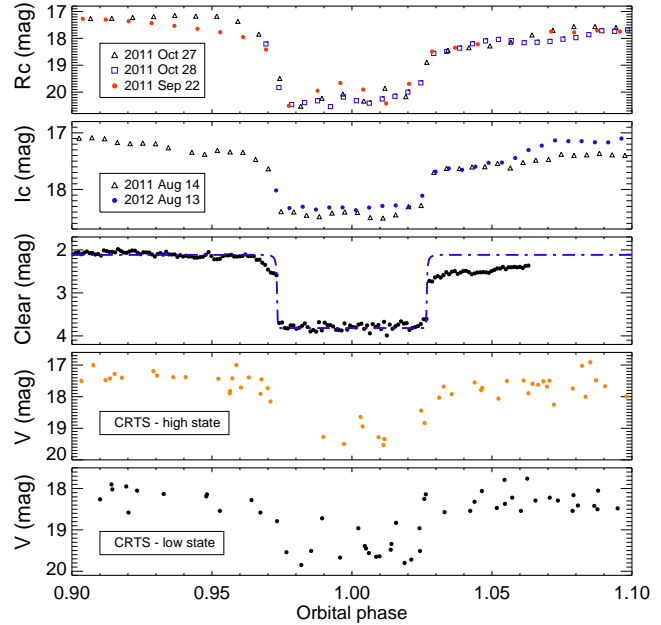}
\caption{Light curves of \ml\, folded with the ephemeris estimated in Section \ref{curvas}. Light curves expanded in the phase interval 0.9-1.1 to detail the eclipse shape. From top to bottom: OPD ${R_C}$ band, OPD ${I_C}$ band, OPD white light, CRTS high state $V$ band and CRTS low state $V$ band. White light magnitudes are not calibrated. The blue dot-dashed line in the white light data (middle panel) is the result of the eclipse fitting discussed in Section~\ref{eclip}.} 
\label{fig:lc2}
\end{figure}

Errors of T$_0$ and of the period are given in brackets. Fig.~\ref{fig:lc2} shows the OPD white light, ${I_C}$ and ${R_C}$ light curves as well as CRTS data, folded with the period and epoch given in the above ephemeris. Despite the poor temporal resolution and the noise, long-spanning CRTS data are very well described by the ephemeris, with the eclipses clearly visible both in the high and low states (see below) with correct phasing. The eclipse has a flat bottom, with a duration of about 15 min. Its ingress is steep and lasts less than 2 min, while the egress is less sharp. The shape of the eclipse is consistent with the eclipsed object being the WD and/or a post-shock region. 

By inspecting the CRTS light curve we note that it can be divided in two distinct brightness levels (high and low) separated by about 1.5~mag, as can be seen in Fig.~\ref{fig:lc3} (top and middle). 
Following \citet{wu2008}, we used a histogram of the magnitudes to choose the limit magnitude between high and low states. Figure ~\ref{fig:lc3} (bottom) shows a clear minimum at 17.85~mag, which we adopted as this limit. However, since \ml~ is an eclipsing system, the classification of a point as low or high can be tricky. 
We performed the classification of datasets of consecutive points (a visit) as described below. The vast majority of the visits spans more than 15 min, which is the eclipses length. Therefore most visits have at least one point out of the eclipse, and that point was used to estimate the brightness state of the corresponding dataset: if at least one point of a visit was brighter than 17.85~mag, then we considered this dataset as belonging to the high state. On the other hand, datasets with out-of-eclipse points fainter than 17.85~mag were classified as in the low state. But some CRTS visits are difficult to classify as high or low state because they include out-of-eclipse points consistent with both states. We tried to use a different limiting magnitude, but it did not help. Some of these points are seen around orbital phases 0.3 and 0.8 in Fig.~\ref{fig:lc3} (middle). A possible explanation is the presence of clouds during CRTS observations.

Fig.~\ref{fig:lc3} (top) shows \ml\ light curve in HJD. There is no typical duration of the high or low states, being the transition between states fast, lasting about a day, which is a common behaviour among polars. According to this classification in distinct brightness levels, all of our OPD photometric data seem to correspond to the high state, except for $I_c$ band from August 14, 2011. 
   
The out-of-eclipse light curves show orbital variation in all bands, but it seems to depend on the brightness state (Figs.~\ref{fig:lc1} and \ref{fig:lc3}). The CRTS data in low state have a modulation of around 1~mag with a minimum around 0.5 orbital phase. The phase dispersion is around 0.5~mag. In high state, the light curve has a larger dispersion, of about 1~mag, and no orbital modulation is observed. Moreover, possible departure from the high state may also be present in OPD data, as can be seen in the ${I_C}$ light curves in Fig.~\ref{fig:lc1}, which were obtained one year apart. We did a careful analysis of objects in the field, and we discarded variation in the comparison star. Flickering with average 0.02~mag amplitude and time-scales of few minutes is also visible in the out-of-eclipse OPD light curves.

\begin{table}
\begin{center}
\caption{Mid-eclipse timings of \ml.%\cvr{centralizei a primeira coluna}
}
\label{tabephe}
\begin{tabular}{c c }
\hline
E          &     T$_{min}$   \\
(cycles)   & (BJD(TDB) 2450000+)  \\
\hline
$-$468   &    5788.750530($\pm$72) \\
$-$221   &    5827.726780($\pm$129) \\
0        &    5862.600061($\pm$118) \\
7        &    5863.704851($\pm$73)\\
1845     &    6153.738822($\pm$75)  \\
4222     &    6528.826641($\pm$16) \\

\hline
\end{tabular}
\end{center}
\end{table}

% ==== POLARIMETRY DISCUSSION ====
%* discuss polarimetric

\subsection{Polarization}
\label{sec_pol}

The polarimetric light curve (Fig.~\ref{pol_p}) confirms \ml\ classification as a polar, since it shows variable polarized emission. These data were probably obtained while it was in the high brightness state because the CRTS data collected near our polarimetric observations are also in high state.

The system presents circular polarization varying in the range $-13$ to 4 per cent in ${R_C}$ band. The circular polarization does not have a standstill at zero, which indicates that a cyclotron component is always contributing to the observed flux in the high state. There are inversions of the polarization sign around phases 0.5 and 0.8. This phenomenon indicates either one emitting region always visible, but observed from the backside in some phases, or two accretion regions. The polarization has its maximum absolute value around phase 0.0, which corresponds approximately to the phase of maximum flux in the out-of eclipse low state light curve. It could be interpreted as if the low state modulation is also mainly due to cyclotron emission from the accretion column. Moreover, both in high and low states, the flux modulation is neither consistent with ellipsoidal variation nor with a heated secondary surface. 

The linear polarization is small, below 5 per cent, and variable. Its maximum values occurs at phases $0.4$ and $0.75$, which are near the phases when circular polarization changes sign: this is consistent with the direction of the emitting region changing from forward to backward. However, our data do not have a good SNR, and these conclusions should be taken with caution.

% ==== SPECTROSCOPY DISCUSSION ====

\subsection{Spectroscopy}

The average spectrum of \ml~ is typical of polars (Fig.~\ref{fig:spec}). It is dominated by the Balmer and \mbox{He\,{\sc ii}} 4686 \r{A} emission features, and also presents several lines of \mbox{He\,{\sc i}} and the Bowen \mbox{C\,{\sc iii}}/\mbox{N\,{\sc iii}} complex at 4630-4660 \r{A}. Emission lines of \mbox{Fe\,{\sc ii}} 5169 \r{A} and \mbox{Si\,{\sc ii}} 6347, 6371 \r{A} are also visible. The continuum is flat with a very slight blue slope and the Balmer lines have inverted decrement which, together with the intense \mbox{He\,{\sc ii}} 4686 \r{A} line, are characteristic of high-ionized optically-thick emission regions. Identification, equivalent width and FWHM of the main emission features in the average spectrum are presented in Table~\ref{tablines}.

The spectrum shows no sign of stellar features. Comparing the flux level to the CRTS data, we see that our spectra of \ml~ were probably obtained in high brightness state. In addition, the absence of cyclotron humps and of secondary features in the spectra is consistent with the system being in a high accretion state, since these features may be diluted by the strong accretion radiation.

\begin{figure}%% l b r t
\centering
\includegraphics[clip,width=0.48\textwidth]{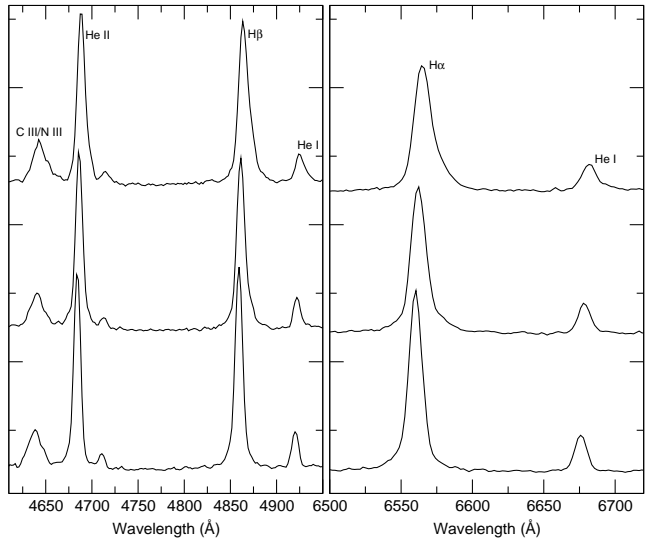}
\caption{Individual spectra of \ml~obtained at orbital phases 0.66, 0.75 and 0.84 (from bottom to top, respectively). The spectra were shifted in the vertical axis for clarity, maintaining relative flux ratios. } 
\label{3specteste}
\end{figure}

Our spectra do not cover a complete orbital cycle. In the analysis of each of the three individual spectra
(Fig.~\ref{3specteste}), the \mbox{He\,{\sc ii}} 4686 \r{A} and Balmer lines have profiles changing from symmetric, at the first spectrum, to profiles with extended red wings in the third spectrum, taken 40 min later. We calculated the radial velocities of the \mbox{C\,{\sc iii}}/\mbox{N\,{\sc iii}} complex, \mbox{He\,{\sc ii}} 4686 \r{A} and Balmer lines, as a function of the orbital ephemeris, by fitting a gaussian function to the peak of the lines (Table~\ref{tabvr}).  The lower limit to the RV amplitude is 300~km~s$^{-1}$. %\cvr{
As these velocities were measured at the line peaks, they should map the velocities of the narrow component seen in polars, probably associated with the mass flux close to the donor star.%}

\begin{table}
\begin{center}
\caption{Identification, EW and FWHM of emission features in the average spectrum of \ml. The FWHM measurements were obtained from Voigt profile fitting.}
\label{tablines}
\begin{tabular}{lcc}
\hline
 Identif.   & $-$EW (\r{A})   &   FWHM (km s$^{-1}$)    \\
\hline

\mbox{He\,{\sc i}} 4388 \r{A}        & 4    &   750  \\ 
\mbox{He\,{\sc i}} 4471 \r{A}       & 16   &  1070   \\ 
\mbox{C\,{\sc iii}}/\mbox{N\,{\sc iii}} 4640 \r{A}  & 14   &  1100   \\ 
\mbox{He\,{\sc ii}} 4686 \r{A}      & 49   &  640   \\ 
H$\beta$         & 70   &  680   \\ 
\mbox{He\,{\sc i}} 4922 \r{A}       & 9    &  550   \\ 
\mbox{He\,{\sc i}} 5016 \r{A}       & 13   &  600   \\ 
\mbox{He\,{\sc ii}} 5412 \r{A}  & 10   &  610   \\ 
\mbox{He\,{\sc i}} 5876 \r{A}   & 21   &  560    \\ 
H$\alpha$        & 80   &  590      \\ 
\mbox{He\,{\sc i}} 6678 \r{A}   & 12   &  540   \\  

\hline
\end{tabular}
\end{center}
\end{table}

\begin{table}
\begin{center}
\caption{Radial velocities of selected lines, in km s$^{-1}$.}
\label{tabvr}
\begin{tabular}{l c c c c }
\hline
phase & \mbox{C\,{\sc iii}}/\mbox{N\,{\sc iii}} 4640 \r{A}	  &  \mbox{He\,{\sc ii}} 4686 \r{A}   &H$_{\beta}$& $H_{\alpha}$   \\
\hline
0.66 &     $-$124 &  $-$106 &  $-$129  & $-$127      	 \\
0.75 & 	     25   &  19     &  $-$12   & $-$30	         \\
0.84 & 	     186  &  169    &  156     & 98	             \\
\hline
\end{tabular}
\end{center}
\end{table}

\section{Cyclotron Models}

We performed a modelling of the cyclotron emission of \ml\ . {\sc Cyclops} code was used to fit the available photometric and polarimetric data and to estimate some geometrical and physical properties of the system. {\sc cyclops} is a radiative transfer code that calculates the cyclotron and bremsstrahlung emission from a 3D post-shock region. The accretion geometry is defined by a bipolar magnetic field, $B_{pole}$, centred in the WD. The plasma temperature ($T$) and density  ($N_e$) structures are defined by shock-like profiles in the radial direction and by an exponential decay or a constant profile in the tangential direction. The shock-like profiles are analytical functions having a shape similar to the magnetised 1D shock solution of \citet{cropper1990} and \citet{saxton2007}. Their parameters are the maximum temperature, the maximum density, and the height of the post-shock region. These functions are presented in equations (1) and (2) of \citet{silva2013}.

Eleven parameters define a model, as listed in the upper part of Table~\ref{tab_cyclops}. The parameters define the geometrical and physical properties of the the post-shock region, which is divided into voxels. The spatial resolution of a model is a free parameter, which defines the voxel size. The emissivities of the four Stokes parameters are calculated according to the $B$, $T$, and $N_e$ of each voxel, and after that, the radiative transfer is calculated, from the bottom to the top of the region, according with the line of sight. The result is an image in each orbital phase for each Stokes parameter. These values are integrated in area to obtain the flux and polarization as a function of the orbital phase. The figure-of-merit to quantify the fitting quality is the $\chi^2$. The genetic algorithm {\sc pikaia} \citep{charbonneau1995} is used to identify the regions in the parameter space having the best models, and then an amoeba based code refines the search. The model can fit several optical bands simultaneously. For a detailed description of the code, see \citet{costa2009} and \citet{silva2013}.

We fitted the high state data of \ml~comprising $V$ light curve from CRTS, R$_c$ light and polarization curves from OPD and I$_c$ light curve obtained 2012 August 13, also from OPD (Fig.~\ref{fig_cyclops}). Magnitudes and polarizations were converted into total and polarized fluxes. The data in each band were binned in 40 phase intervals. Although the data are not simultaneous, we consider that the brightness states are similar. The large error bars of the $V$ light curve are probably the result of a combination of measurements taken in slightly different brightness states. Photometric measurements obtained near eclipse, between phases 0.95 and 0.1, were analysed separately (see Section \ref{eclip}), as {\sc cyclops} does not handle the eclipse of the post-shock region by the secondary.

In our minimisation procedure, we keep free all parameters listed in Table 5 except for the inclination, which was limited between 75 and 90 deg due to the presence of total eclipse. In the modelling, we used the following frequencies to represent each optical band: $V$ = 5.45$\times 10^{14}$~Hz, $R_C$ = 4.49$\times10^{14}$~Hz and $I_C$ = 3.8$\times10^{14}$~Hz. The results of our modelling are presented in the next section.

\subsection{Modelling discussion}
\label{sec_model_discussion}

As already discussed in Section~\ref{sec_pol}, the inversion of the polarization sign may be observed in two situations: (i) two emitting regions; (ii) one emitting region whose magnetic field lines alternate direction back and forth along the orbital cycle. As we found a reasonable fit using only one region, we adopted this solution, therefore minimising the number of free parameters of the model.

\begin{figure*}%% l b r t
\centering
\includegraphics[trim=3.0cm 1cm 4cm 1.1cm, clip,width=0.50\textwidth]{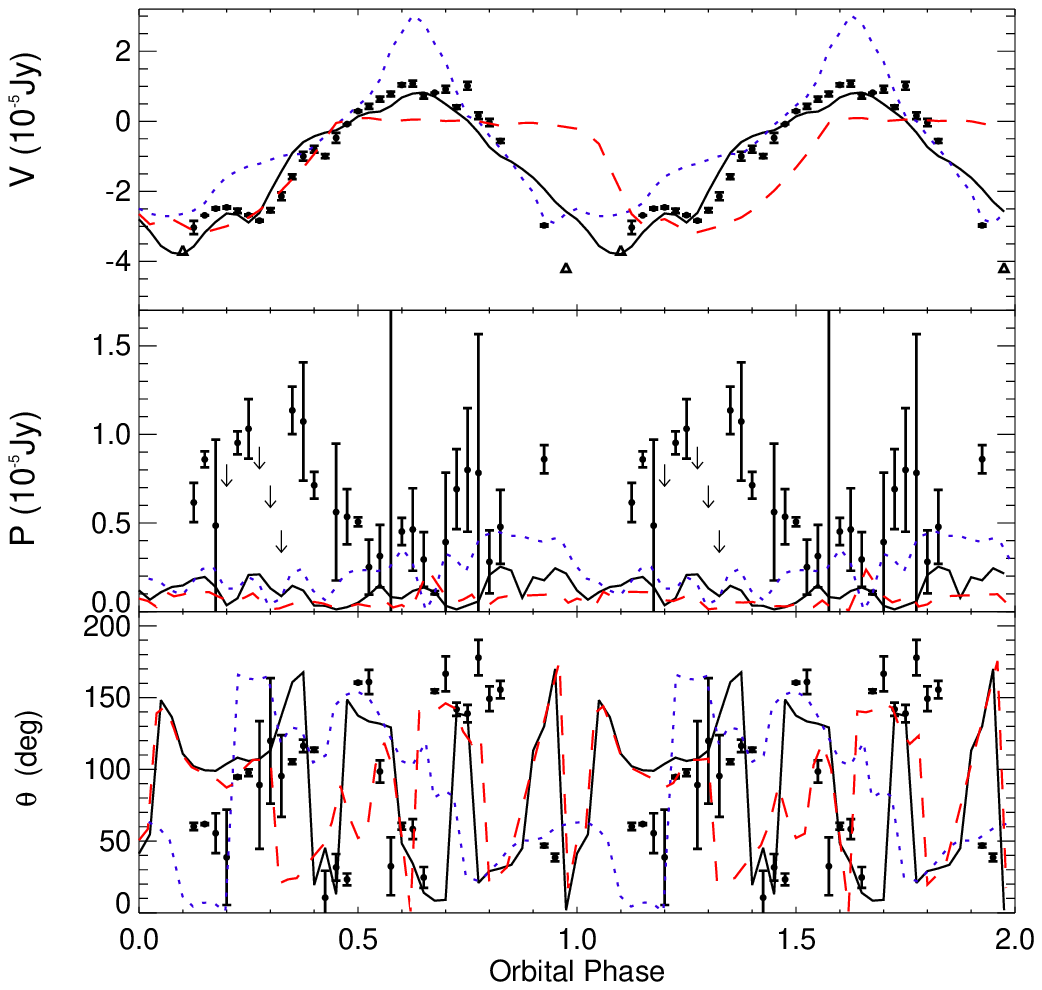}
\includegraphics[trim=3.5cm 1cm 4cm 1cm, clip,width=0.48\textwidth]{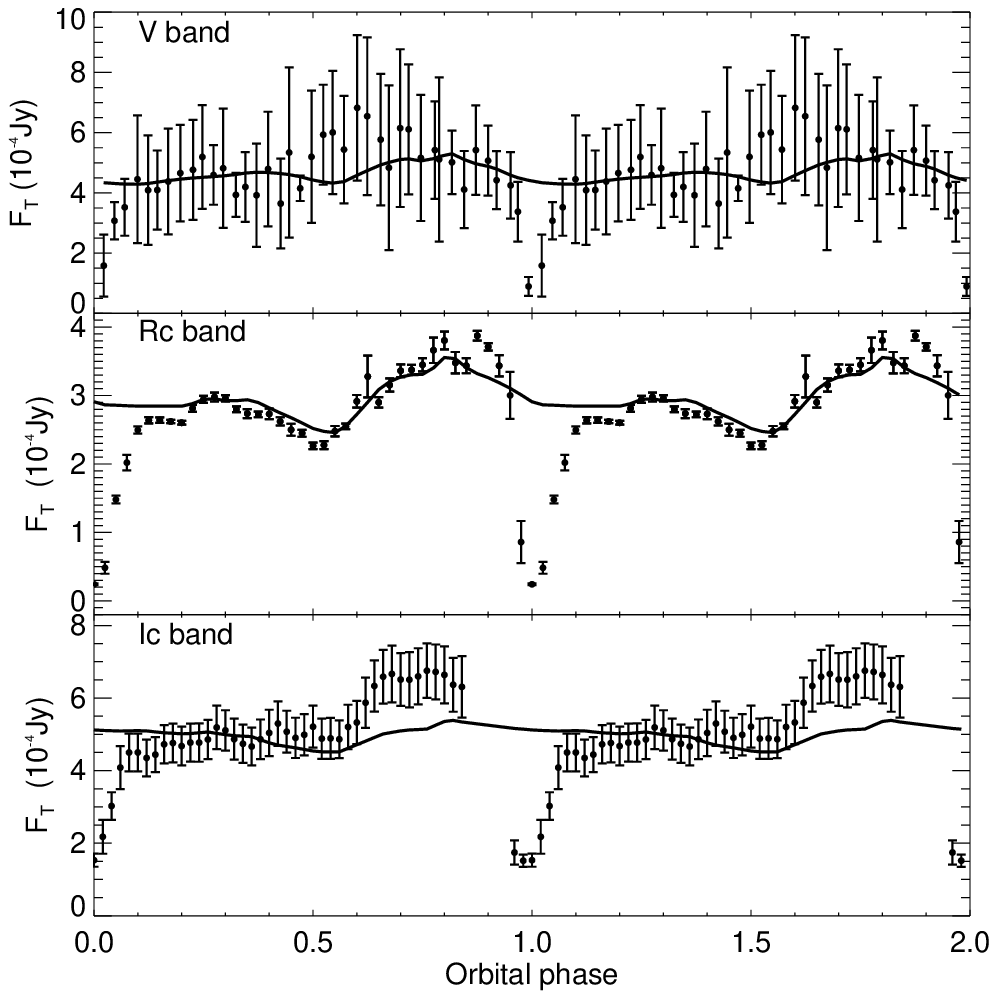}
\caption{\ml~ data (points and triangles- the last with no erros estimates) and the best-fit model (lines) described in text and whose parameters are presented in Table~\ref{tab_cyclops}. Left panel: Polarization modelling. From top to bottom, the panels show circular polarized flux ($F_V$), linear polarized flux ($F_P$) and angle of linear polarization ($\theta$). The black dots are the $R_C$ polarization data and the black solid line represents the best $R_C$ model. The blue dotted line is the $V$ band model prediction and the red dashed line represents the model prediction in $I_C$ band. Right panel: Total flux modelling ($F_T$). From top to bottom, the panels shows the light curves in $V$, ${R_C}$ and ${I_C}$ bands. The black lines are the best models. } 
\label{fig_cyclops}
\end{figure*}

Using {\sc pikaia} with distinct parameters intervals, the best fits were always found in a same group of models. The fits considering layers with constant values of $N_e$ and $T$ are worse than those in which these quantities are let to vary in the direction perpendicular to the column height. The model having the smallest $\chi^2$ is shown in Fig.~\ref{fig_cyclops} and its parameters are presented in Table~\ref{tab_cyclops}. We discuss this model below.

\begin{table}
\caption{Physical and geometrical properties estimated for MLS110213.} 
\label{tab_cyclops}
\begin{center}
\begin{tabular}{ll} 
\hline
 Cyclops input parameters                         & Fitted values \\
\hline 
$i$                                               &  77$\degr$  \\
$\beta$                                           &  39$\degr$  \\
$\Delta_{long}$                                   &  87$\degr$ \\
$\Delta_R$                                        &  0.29  \\
$h$                                               &  0.18 $R_{WD}$ \\
$f_l$                                             &  0.29   \\
$B_{pole}$                                        &  35 MG \\
$B_{lat}$                                         &  65$\degr$ \\
$B_{long}$                                        &  49$\degr$ \\
$T_{max}$                                         &  29 keV \\
$N_{e~max}$                                        & 14.3 cm$^{-3}$     \\
\hline
Model results                                     & Values \\
\hline
$B_{reg}$                                         & 18 -- 33 MG\\
$\langle T \rangle$                               & 9 keV    \\
$T_{pond}$                                        & 5 keV      \\
$T_{range}$                                       &  1-29 keV \\
$\delta_{phase}$                                  & $-0.11$ \\
$\chi^2$                                          & 0.395  \\
\hline 
Stellar parameters                                   & Values    \\
\hline
$q$                                               & 0.42 \\
M$_1$                                             & 0.67 M$_{\odot}$ \\
R$_1$                                             & 0.012  R$_{\odot}$ \\
M$_2$                                             & 0.28 M$_{\odot}$ \\
R$_2$                                             & 0.36 R$_{\odot}$ \\
\hline

\end{tabular} 
\end{center}

Parameters description - $i$: orbital inclination; $\beta$: colatitude of the reference point of the coordinate system inside the emitting region; $\Delta_{long}$: one-half of the azimuth of the threading region; $\Delta_R$:  one-half of the radial extension of the threading region; $h$: height of the post-shock region; $f_l$: position in the longitudinal direction of the threading point relative to the geometrical centre of the threading region; $B_{pole}$: polar magnetic field intensity; $B_{lat}$: latitude of the magnetic axis; $B_{long}$: longitude of the magnetic axis; $T_{max}$: maximum electronic temperature; $N_{e~max}$: log of maximum electronic density; $B_{reg}$: magnetic field in the post-shock region; $\langle T \rangle$: mean eletronic temperature; $T_{pond}$: mean temperature weighted by the square density ($\sum(T_{vox} \ N_{vox}^{2})/ \sum N_{vox}^{2}$; $T_{range}$: range of temperatures of in the post-shock region; $\delta_{phase}$: phase shift applied to the model. 

\end{table}

The model reproduces quite well the $R_C$ circular polarization (Fig.~\ref{fig_cyclops}, left/top panel, black lines). In particular, the observed sign inversion is well fit. The observed linear polarization flux is small and noisy, but larger than that produced by the model. 

The flux modulation in $V$, $R_C$, and $I_C$ bands are reasonable well fit (Fig.~\ref{fig_cyclops}, right panel). Figure ~\ref{fig_cyclops} also presents the polarization predictions to $V$ and $R_C$ bands (blue dotted and red dashed line, respectively). The predicted circular polarization in $V$ band is similar in shape, but has a larger amplitude than $R_C$ band, and the sign inversion is more pronounced. $I_c$  band model prediction shows null circular polarization from phase 0.5 to 1.1.

\begin{figure}%% l b r t
\centering
\includegraphics[trim= 4.4cm 16cm 2cm 4cm,clip,width=0.48\textwidth]{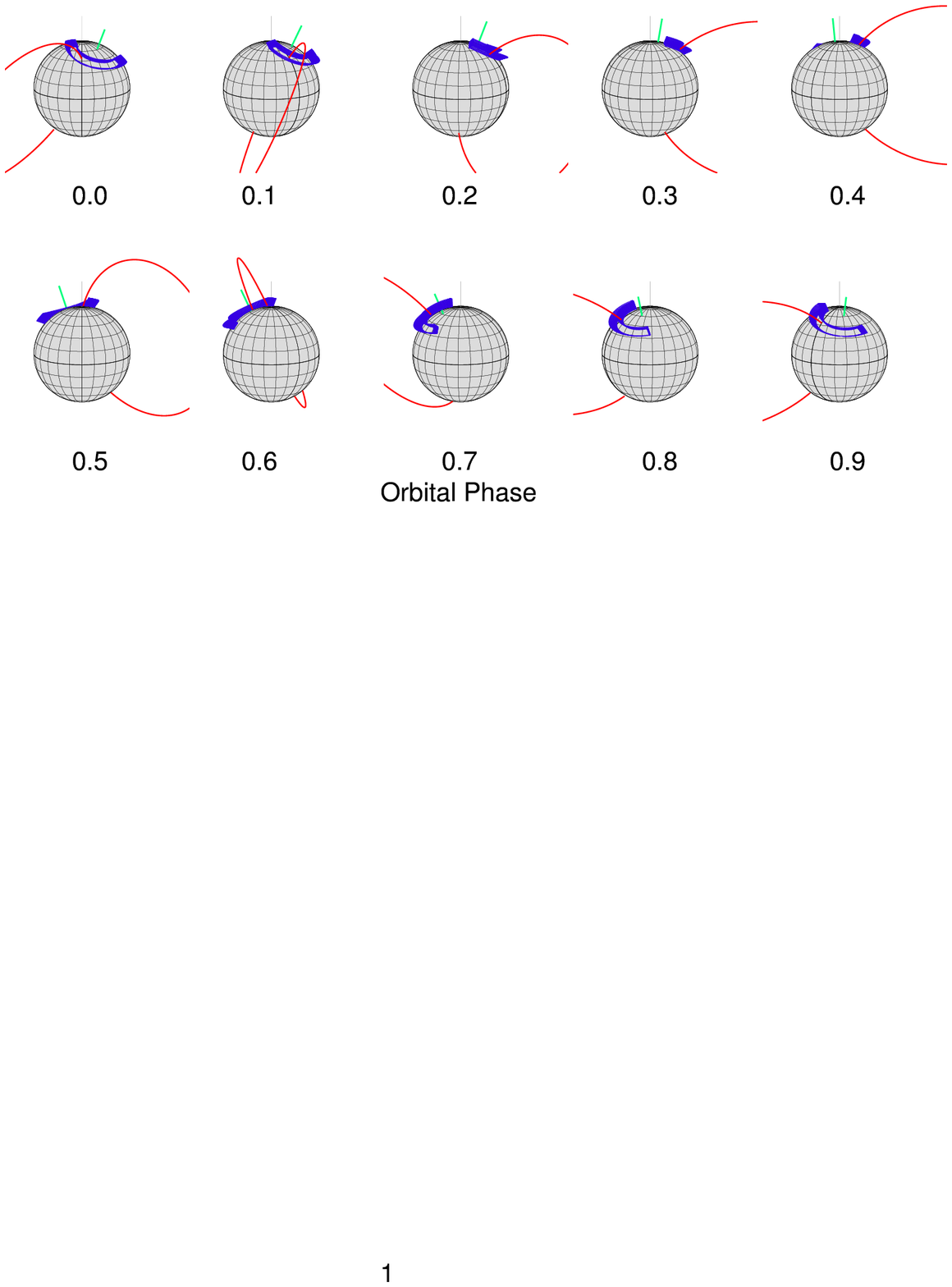}
\caption{Geometrical representation of the emitting region on the WD surface in \ml. The post-shock region is represented by its walls (blue lines). The curved red line near the centre of the emitting region is a magnetic field line threading the emitting region. It is shown to represent the accretion column geometry. The radial green line is the magnetic axis.} \label{st_seed_multi}
\end{figure}

Figure \ref{st_seed_multi} is a geometrical cartoon of our modelling of the emitting region on the WD surface along the orbital cycle. From phase 0.7-0.1 the emitting region is visualised from the top, when the circular polarization is negative and larger. From 0.2 to 0.6 the region is partially occulted and is seen from the opposite direction, hence the circular polarization is positive and smaller. The sign inversions can be associated with the following configurations: the first sign inversion occurs around phase 0.5, when the region is seen completely by the opposite side for the first time; the second inversion occurs around phase 0.8, when the entire region starts to point to the observer again.  Therefore, we can conclude that the circular polarization sign inversion can be due the changes in the angle of view of one dominant emitting region. The change in the signal does not happen early in phases 0.2 or 0.3 because of the partial self-eclipse. The self-eclipse could also decrease the value of the absolute circular polarization flux. 

The best fit model corresponds to an extended region, with a relatively high magnetic latitude and moderate electronic temperature. Orbital inclination from 75$\degr$ to 81$\degr$ produces very similar $\chi^2$ values: inclinations out of the range 74$\degr$ -- 85$\degr$ increase $\chi^2$ by at least 30\%. The best-fit magnetic field in the magnetic pole, $B_{pole}$, is 35~MG. This corresponds to a magnetic field in the region varying from 18 to 33~MG. Values of $B_{pole}$ between 34 and 36~MG produce very similar $\chi^2$ values, while $B_{pole}$ of 30 and 40~MG produce $\chi^2$ 30\% larger. The region mean temperature is 9~keV. The mean temperature weighted with the square density, $T_{pond}$, is around 5~keV.
For the temperature and density we found that the ranges 24 -- 30 K and 14 -- 14.3 $cm^{-3}$  provide small changes in the $\chi^2$. Out of the limits of 22 -- 35 K and 13.8 -- 14.6 $cm^{-3}$, the $\chi^2$ value increases by at least 30\%.
These are low values of temperature for a polar post-shock region and indicate a dominance of soft emission in X-ray. The maximum temperature in the emitting region, T$_{max}$, corresponds to the temperature in the shock front. Using the approach of \citet{1997ApJ...474..774F} and the \citet{nauenberg1972} mass-radius relation, we estimated a mass of 0.67~M$_\odot$ for the WD from T$_{max}$. The orbital period of \ml~ translates to a secondary mass, M$_2$, of 0.279~M$_\odot$ \citep{knigge2011}, resulting in a mass ratio, q, of 0.42. The stellar parameters are also summarised in Table 5.

The presented modelling of the accretion column can be refined using new additional data, for instance, observations in other broad bands, simultaneous data acquisition and/or spectropolarimetry is desirable.

\section{Eclipse fitting}
\label{eclip}
We fitted the eclipse profile seen in the clear band observations considering a simple model of two spherical stars in a circular orbit. The clear data present the best temporal resolution covering the eclipse. The eclipse is box shaped indicating a total eclipse of a very small object.

We determined M$_2$=0.279 M$_{\odot}$ and R$_2$= 0.37 R$_{\odot}$ by interpolating the semi-empirical CV donor star sequence \citep{knigge2011} as a function of the orbital period of the system. The flux of the donor star is considered negligible and M$_{1}$ is obtained from the mass ratio, $q$ = M$_{2}$/M$_{1}$. We use the masses to determine the distance between the stars from the Kepler Third Law and use \citet{nauenberg1972} relationship to determine R$_{1}$ from a given M$_{1}$. Thus the parameters of the eclipse fitting are the orbital inclination $i$ and the mass ratio $q$.

There is a degeneracy in the possible values of $i$ and $q$ that can fit the eclipse profile. M$_{1}$ will reach the limit of 1.4 M$_{\odot}$ at $q$ = 0.20 and $i$ = 79$\degr$. We can increase $q$ while decreasing $i$ until it reaches the minimum inclination limit for a total eclipse at $i$ = 73$\degr$ and $q$ = 0.75. Therefore, the observed eclipse width can be reproduced using the following intervals of parameters:  M$_1$ = 0.37 -- 1.4 M$_{\odot}$, with $q$ = 0.75 -- 0.20 and $i$ = 73$\degr$ -- 79$\degr$.

The {\sc cyclops} modelling shows no significant variation in the fitting for the interval $i$ = 75$\degr$ -- 81$\degr$, indicating that the {\sc cyclops} geometry can also describe the observed eclipse profile. The blue dot-dashed line in the middle panel of Figure \ref{fig:lc2} shows the eclipse fitting using $q$ = 0.42, as estimated in the previous section. This value corresponds to M$_1$= 0.67 M$_{\odot}$ and R$_1$ = 0.012 R$_{\odot}$, keeping M$_2$ constant. The inclination that fits the eclipse profile is $i$ = 75.8$\degr$, in agreement with {\sc cyclops} estimated interval.  

In polars, the emitting region can be much brighter than the WD itself, therefore determining the eclipse profile. The \ml\ eclipsed region is much smaller than the secondary star, as shown by the fast ingress and egress of the total eclipse. But the data do not have enough time resolution to allow the determination of the size of the eclipsed region.

We consider that the approximation of a spherical secondary star is valid because, at conjunction, the projected area of the secondary star on the plane of sky is minimum. To check this assumption, we compare our results with those of \citet{chanan1976}, who consider the actual geometry of a Roche-lobe-filling companion and a pointlike eclipsed object. We measured the half-eclipse width of \ml\ as 9.36$\degr$. Accordingly with Table 1 of \citet{chanan1976}, the mass ratio of q=0.42 results in an inclination of $i$ =  76.7$\degr$, in agreement with our result. Moreover, \citet{andronov2014} compare the use of a spherical approximation and an elliptical projection onto the celestial sphere for a secondary that fills its Roche lobe and found that, in the spherical approximation, the inclination is underestimated by about 2$\degr$, a value which is inside our uncertainties.

\section{Distance estimation}
\label{dist}

In order to estimate the distance to \ml, we adopted the Period-Luminosity-Colours (PLCs) method derived by \citet{ozdonmez2015}
The method uses the orbital period of the system together with J, K$_s$ and W1 magnitudes, from 2MASS \citep{skrutskie2006} and WISE \citep{wright2010}, to estimate absolute magnitudes of CVs. The method is based on the \citet{1976MNRAS.174..489B} relation which states that, for late-type main sequence stars, the surface brightness is nearly constant in the near-infrared. The PLCs relation is calibrated by the trigonometric parallaxes of 25 CVs. The accuracy in M$_J$ provided by the PLCs method is $\pm0.29$ mag. As the PLCs method does not consider other sources of infrared light except the secondary star, it yields a lower limit to the distance. The PLCs relation is:

\begin{equation}
 M_J = 5.966~-~4.781 \log P~+~5.037 (J-K_S)_0~+~0.617 (K_S-W1)_0
\label{eq_plc}
\end{equation}

\noindent where $P$ is the orbital period given in hours and the de-reddened colours $(J-K_S)_0$ and $(K_S-W1)_0$ are calculated from $J_0~=~J~-~A_J$, $K_{S0} = K_S-A_{K_S}$ and $W1_0=W1-A_{W1}$. 
We adopted $A_J=0.887 \times E(B-V)$, $A_{K_S}=0.382 \times E(B-V)$ and $A_{W1}=0.158 \times E(B-V)$ \citep{ozdonmez2015}. 

The infrared magnitudes of \ml, obtained from the 2MASS and WISE database, are $J=15.928$,  $H=15.264$, ${K_S}~=~15.000$, $W1=14.680$ and $W2=14.163$. The colour excess in the direction of \ml, obtained from \citet{2011ApJ...737..103S} maps, is $E(B-V)={0.0941}$. With those values and considering the accuracy in M$_J$, we estimate a distance to \ml\ of $406\pm54$ pc, assuming that the secondary is the only infrared light source in the system. Despite being a lower limit, this distance should not be very different from the true value, since our {\sc cyclops} modelling indicates that the cyclotron contribution to the total infrared flux of the binary system is small. This may not be the case for other magnetic CVs. In non-magnetic CVs, the infrared flux may have contributions from other structures like the outer regions of the accretion disc.

\section{Discussion and conclusions}

We have obtained photometric, polarimetric and spectroscopic data of \ml, which is one of the polar candidates identified and monitored by the Catalina Real Time Transient Survey. We confirm the system as a new eclipsing polar.

The optical spectrum, obtained in the 0.7 - 0.88 phase interval, is very similar to the spectra of other polars. It shows a flat continuum and an inverse Balmer decrement, and also conspicuous features of highly ionized optically thick regions, like strong emission lines of \mbox{He\,{\sc ii}} 4686 \r{A} and \mbox{C\,{\sc iii}}/\mbox{N\,{\sc iii}} 4640 \r{A}. There is no clear evidence of the secondary star in our spectrum. 

Photometric data obtained at CRTS and OPD show that \ml~ presents deep (3 mag) eclipses with flat bottoms, indicating that the system has high orbital inclination. From OPD mid-eclipse timings we estimated an orbital period of 3.787~h, which places \ml~ as the $5^{th}$ eclipsing polar found above the orbital period gap.

The long term observations obtained by CRTS clearly show the system in two distinct brightness states, separated by 1.5 mag. In the high brightness state the out-of-eclipse light curve is almost flat, while in low state a strong modulation is observed, with a maximum at phase 0.175. The presence of distinct brightness states is typical of polars and are understood as variations at the mass accretion rate. 

\ml~ presents circular polarization in the $R_c$ band (obtained in high state) varying from -13 to 4 per cent with a maximum around phases 0 - 0.125, consistent with cyclotron emission from the accretion column. The polarization does not have a standstill at zero, and there are phases when the signal changes, which may indicate either one emitting region always visible, but observed from the backside in some phases, or two accretion regions.

The {\sc cyclops} modelling of high state V, $R_c$ and $I_c$ light curves and of one $R_c$ polarimetric curve provided an estimate of the system parameters and a description of the main observed features, including the eclipse profile. The model agrees with the interpretation that \ml~ has one main circumpolar emitting region, although the possibility that it could be a two-pole accretor cannot yet be ruled out. This model constrains the magnetic field from 30 to 40 MG and the orbital inclination from 75$\degr$ to 85$\degr$. The eclipse fitting corroborates {\sc cyclops} results. Nevertheless, additional polarimetric data in V and $I_c$ band are necessary to a better determination of the system parameters.

The field of \ml\ was observed by ROSAT, RXTE, Integral and Swift observatories, but the source was not detected by none of them.
Using data from the JEM-X camera at Integral, we estimated an upper limit of 1.3$ \times 10^{-10}$~erg~cm$^{-2}$~s$^{-1}$ for the X-ray flux in the energy range of 2--5~keV. This flux is quite high, and we consider that the Integral non-detection is due to a short integration time. The ROSAT Bright Survey detection limit flux is 2$\times10^{-12}$~erg~cm$^{-2}$~s$^{-1}$ in the 0.12--2.48~keV band, considering a typical model for X-ray polars emission \citep{pretorius2013}. Adopting this flux limit and the distance estimate of $406$~pc, the X-ray luminosity of \ml~ is lower than 3.9$\times 10^{-31}$~erg~s$^{-1}$. All ROSAT polars with X-ray luminosities similar or smaller than this value are nearer systems \citep{pretorius2013}. Therefore the low X-ray flux of \ml~ is possibly explained by the distance to the source and accentuated by the low temperature of the emitting region, as estimated by the {\sc cyclops} fit. This fit predicts a modulation of the X-ray light curves with the minimum flux from phases 0.2 to 0.5, caused by partial self-eclipse of the emitting region. Deep X-ray observations of \ml\ would be very useful to constrain the geometric and physical properties of the system.
 
\ml\ is among the few polars not found by X-ray surveys. Some of them are classified as low accretion-rate polar (LARPs), a class now also recognised as pre-polars \citep[e.g.,][] {schwope2002,webbink2005,schmidt2005,breedt2012}. \ml\ shares the low X-ray flux and low temperature of this class, but the low X-ray brightness may be explained by the \ml\ distance. Its post-shock region seems to be dense, produced by a Roche lobe overflow mass transfer. This diverges from \ml\ being a pre-polar, since in them the mass transfer is supposed to originate from a secondary wind, with mass transfer rates about 100 times smaller than in standard polars. The most important signature of a low mass transfer rate, the cyclotron harmonics, is also lacking in the \ml\ spectra. Therefore, we do not classify \ml\ as a LARP/pre-polar.

\citet[and a series of previous articles]{li1998} study the effect of the WD magnetic field on the secondary magnetic braking. They show that the interaction of the magnetic fields of both stars can decrease the magnetic braking. This effect is more pronounced if the magnetic dipole axis is parallel to the WD spin axis. Hence they concluded that the evolution of magnetic CVs should take into account the direction of magnetic dipole axis. They argued that most polars tend to have magnetic field dipolar axis approximately aligned with the WD spin axis (at angles smaller than 25$\degr$) and, in this situation, the MB can be lessened by at least an order of magnitude in comparison with non-magnetic CVs. \citet{webbink2002} shows that this decrease in the magnetic braking can populate the period gap in polars. The inclination of the WD dipolar magnetic field obtained by our {\sc cyclops} modelling of \ml\ is 25$\degr$. This value is consistent with the "aligned" field scenario.

\ml\ is a new eclipsing polar above the period gap. Its post-shock region is cooler than in most mCVs, indicating a low-mass WD. \ml\ populates a rare sub-group of polars, near the upper limit of the period distribution, important to understand the evolution of mCVs. Our modelling is a first step to determine parameters as the magnetic field intensity and inclination, which could be used to constrain evolutionary models of mCVs.

\section*{Acknowledgements}
	
KMGS acknowledge Roland Walter for helping with the X-ray data. We acknowledge Andrew Drake for kindly providing us with CRTS data before the public release. The CRTS survey is supported by the U.S.~National Science Foundation under grants AST-0909182. The CSS survey is funded by the National Aeronautics and Space Administration under Grant No. NNG05GF22G issued through the Science Mission Directorate Near-Earth Objects Observations Program. This publication makes use of data products from the Two Micron All Sky Survey, which is a joint project of the University of Massachusetts and the Infrared Processing and Analysis Center/California Institute of Technology, funded by the National Aeronautics and Space Administration and the National Science Foundation. This publication makes use of data products from the Wide-field Infrared Survey Explorer, which is a joint project of the University of California, Los Angeles, and the Jet Propulsion Laboratory/California Institute of Technology, funded by the National Aeronautics and Space Administration. This study was partially supported by: CNPq (CVR:306103/2012-5, KMGS:302071/2013-0) and FAPESP (LAA: 2012/09716-6 and 2013/18245-0; CVR: 2013/26258-4). We thank the referee for the valuable suggestions.

\end{document}